# Long Time Stability of the Energy Scale Calibration of a Quantum 2000

UWE SCHEITHAUER

*SIEMENS AG, CT T DE HW1, Otto-Hahn-Ring 6, 81739 München, Germany*

Phone: + 49 89 636 – 44143
E-mail: uwe.scheithauer@siemens.com, scht.uhg@googlemail.com

## Abstract

According to the international standard ISO 15472 the energy scale of an XPS instrument, type Physical Electronics Quantum 2000, was calibrated. It is shown, how the procedures of the ISO 15472 were adapted to the hardware and software design of the Quantum 2000. The long time stability of the energy scale calibration of the XPS instrument was investigated. The instrumented was operated with a satisfying energy scale calibration over a period of 8 years. All the time energy differences between certain peaks could be measured with the chosen precision of the energy scale calibration.

## Introduction

An X-ray photoelectron spectrometer measures the kinetic energy of electrons emitted after sample excitation using an X-ray source. For the external photoelectric effect it is the binding energy of a certain atomic level. For an Auger process it is the energy difference of three atomic levels. Additional energy correction terms describe the properties of solid state samples, for instance. Each photo or Auger electron has a true peak energy due to the inherent physical properties of an element and the physical and chemical interactions of an atom with its surroundings.

The aim of the energy scale calibration of an XPS instrument is to keep the energy deviation ΔE between the true peak energy and the measured peak energy within predefined limits and to guarantee the long time stability of the energy scale calibration. Thanks to the efforts of many people ***[1-14]*** the reference energies of certain Au, Ag and Cu peaks are defined and procedures of energy scale calibration are elaborated. Based on these results, as well the reference values for the peak positions on the binding energy scale for the Au4f7/2, Ag3d5/2, Cu $L_3$VV and Cu2p3/2 peak, stimulated by Mgk$_\alpha$, Alk$_\alpha$, or monochromatic Alk$_\alpha$ X-rays, as the procedure how to calibrate an XPS instrument are described by the international standard ISO 15472 ***[15;16]***.

For a modern XPS instrument with a primary focused X-ray beam ΔE may be a function of a wide variety of instrumental operation parameters. These parameters are the diameter of the primary X-ray beam, the pass energy of the energy analyser or the lateral distance of the beam relative to the origin on the dispersive axis of the monochromator, for instance. Due to the time effort needed, the energy scale calibration was not applied to all possible operation parameter combinations of a Quantum 2000. The parameters were restricted to a reasonable set.

This study documents the initialization of energy scale calibration and the long time stability of the energy scale calibration of a Physical Electronics Quantum 2000 X-ray microprobe, instrument no. 78, delivered in 2001. The results obtained over a period of 8 years are discussed.

# Instrumentation

The Quantum 2000 is an XPS instrument with a focused primary X-ray beam. The spatial resolution of a Quantum 2000 XPS microprobe is achieved by the combination of a fine-focused electron beam generating the X-rays on a water cooled Al anode and an elliptical double-focusing mirror quartz monochromator ***[17-19]***, which monochromatizes and refocuses the X-rays to the sample surface. This way, the X-ray beam scans across the sample as the electron beam is scanned across the Al anode. On the sample surface an area of 1.4 x 1.4 $mm^2$ can be scanned by applying electronic deflection voltages to the electron beam. Fig. 1 shows a schematic drawing of the principal components of a Quantum 2000 X-ray microprobe. By controlling the electron beam diameter, nominal X-ray beam diameters between approximately 10 µm and 200 µm are adjustable. Details of the quantitative lateral resolution are discussed elsewhere ***[20]***. The X-ray beam power is proportional to the X-ray beam size. Using a rastered X-ray beam, sample features are localized by X-ray beam induced secondary electron images.

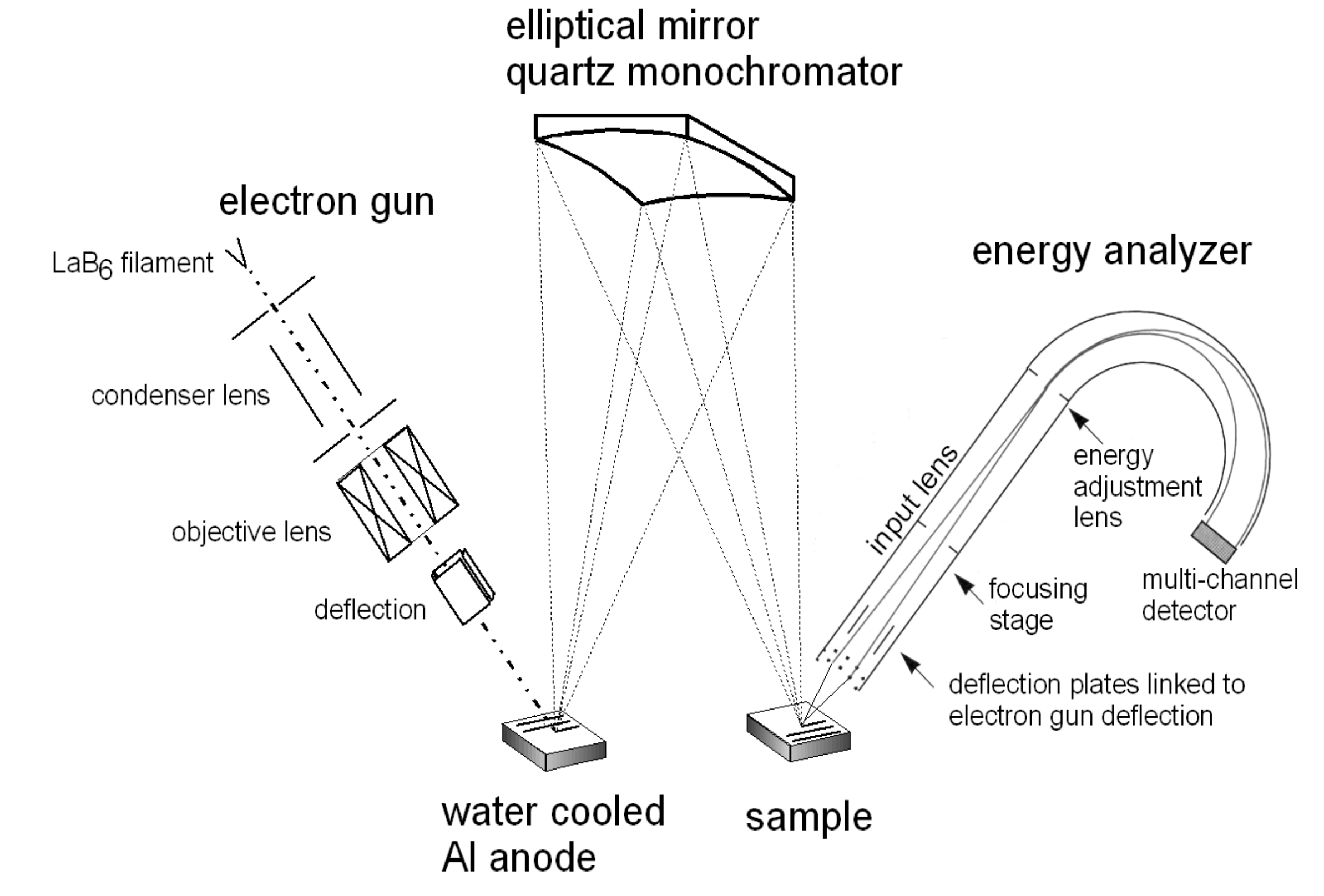

Fig. 1: schematic drawing of the principal components of a Quantum 2000 X-ray microprobe

| **X-ray beam** | | **energy analyser** | |
|---|---|---|---|
| diameter [µm] | power [W] | pass energy [eV] | binding energy grid point intervals [eV] |
| 5 | 1.25 | 2.95 | 0.025 / 0.05 / 0.1 / 0.2 |
| 10 | 2.5 | 5.85 | 0.025 / 0.05 / 0.1 / 0.2 |
| **20** | 5 | **11.75** | **0.025** / 0.05 / 0.1 / 0.2 |
| 50 | 12.5 | 23.5 | 0.025 / 0.05 / 0.1 / 0.2 |
| 100 | 25 | 29.95 | 0.125 / 0.25 / 0.5 / 1.0 |
| 200 | 50 | **46.95** | **0.200** / 0.40 / 0.8 / 1.6 |
| high power (rastered on Al anode) | 100 | 58.70 | 0.125 / 0.25 / 0.5 / 1.0 |
| | | **93.9** | **0.200** / 0.40 / 0.8 / 1.6 |
| | | 117.4 | 0.125 / 0.25 / 0.5 / 1.0 |
| | | 187.85 | 0.200 / 0.40 / 0.8 / 1.6 |

Tab. 1: selectable operation parameters of the X-ray beam and the energy analyser
settings used for the energy calibrations are highlighted.

By emittance matching the analyser acceptance area is synchronized with the X-ray beam position on the sample. Voltages which are synchronized with the raster of the exciting electron beam are applied to electrostatic deflection plates at the analyser entrance for this purpose. Dynamic dispersion compensation is used to compensate the energy shifts due to energy variations of the primary X-rays beam while the beam position is shifted along the disperse direction of the monochromator. The retarding potential of the analyser input lens is varied according to the disperse direction raster position of the X-ray beam ***[19]***.

To enhance the instruments sensitivity, the energy analyser is equipped with a multichannel electron detector. It uses 16 discrete channels for parallel detection. The geometrical lateral distance of these separate detector channels defines the possible combination of the energy analyser pass energy settings and binding energy grid point intervals (see tab. 1).

For flat mounted samples as used here in a Quantum 2000 the incoming X-rays are parallel to the surface normal. In this geometrical situation, the geometrical energy analyser take off axis and the differentially pumped $Ar^+$ ion gun, which is used for sputter cleaning and depth profiling of the samples, are oriented ~45° relative to the sample surface normal.

The instrument operates in an air-conditioned temperature stabilized laboratory at our site. To avoid thermal drifts of the electronics all components are continuously operated. Electrical components are only powered off for service issues.

# Instrumental Operation Parameters used for Energy Scale Calibration

The Quantum 2000 is operable using a large variety of instrumental parameters. The selectable settings of the X-ray beam and the energy analyser are summarized in tab. 1. To make the energy scale calibration practicable, one has to restrict the operation parameters to a reasonable set. All energy calibration data were measured with the 20 µm X-ray beam. Using this beam size, on the one hand the X-ray intensity and thereby the count rates are sufficiently high to avoid long measurement times. On the other hand, it is small enough to avoid energy broadening of the X-ray energy. Energy broadening is due to variations of the X-ray impact angle relative to the disperse direction of the monochromator. And these variations are larger for X-ray beams having bigger beam sizes. The energy analyser pass energy settings were chosen with respect to the count rates (see fig. 2) and the full width at half maximum (= FWHM) of the peaks (see fig.3). The pass energy of 93.9 eV is suitable for high count rate survey spectrum measurements resulting in good detection limits for low concentration contaminations. For high energy resolution measurements of peak positions the 11.75 eV setting is used. And the 46.95 eV setting is an intermediate one.

All measurements were done in the origin of the coordinate system for the XPS instrument. This way neither deflection voltages changes for the emittance matching nor additional retarding voltage variations for compensation of a primary X-ray energy shift were used.

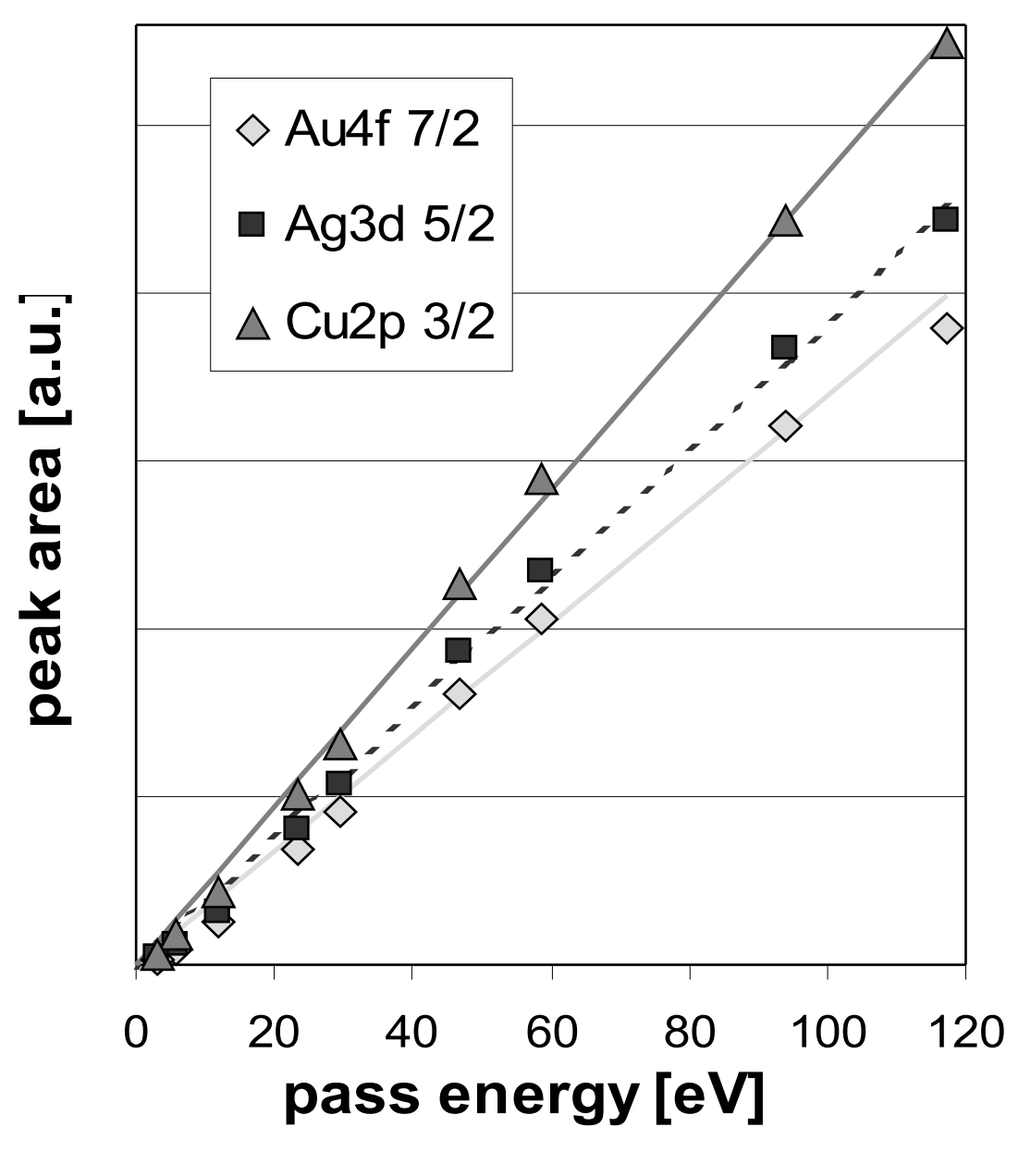

Fig. 2: count rate as function of pass energy

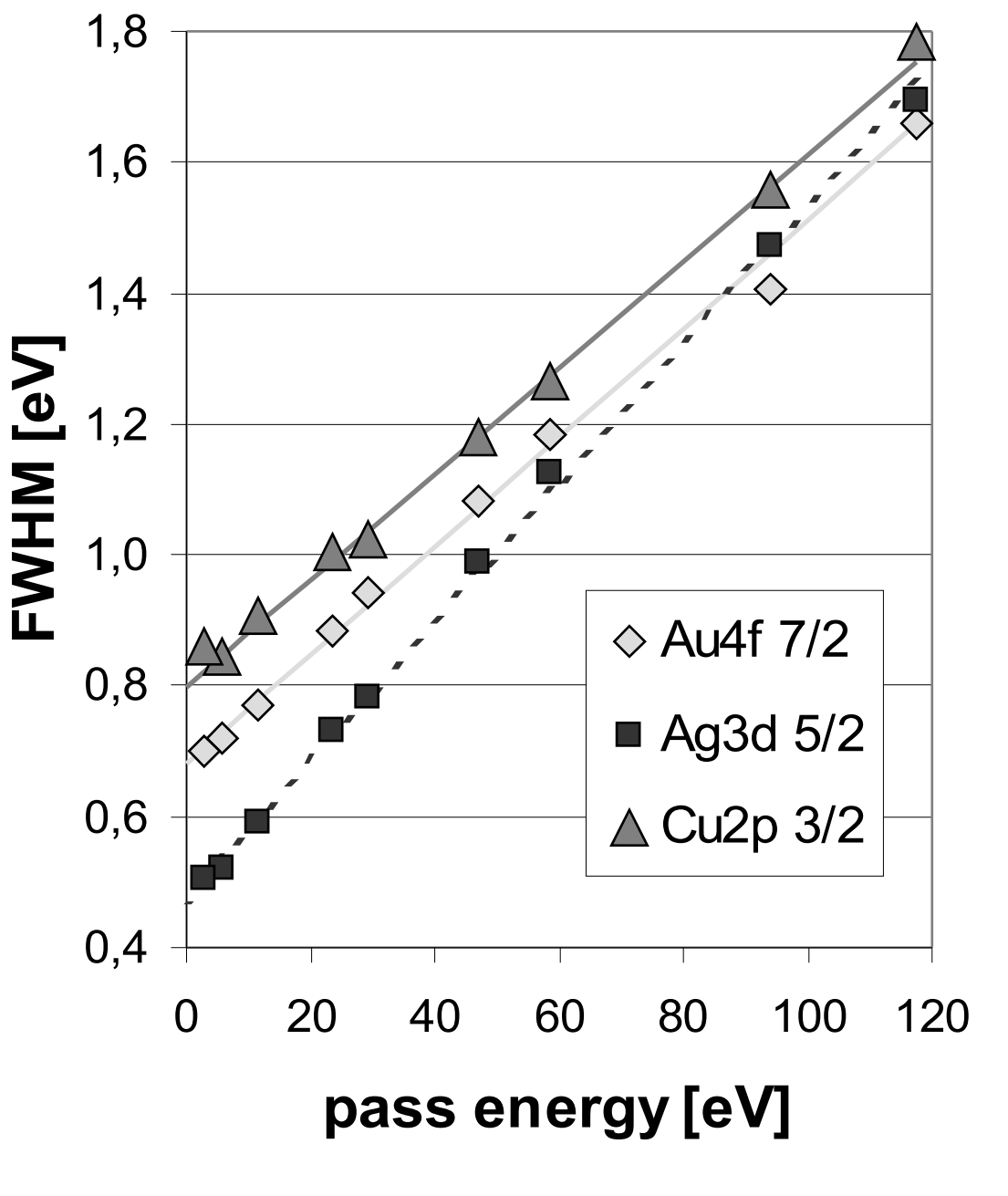

Fig. 3: FWHM as function of pass energy

# ISO 15472 Guidelines and Procedures

The standard ISO 15472 ***[15]*** defines how to calibrate the energy scale of an XPS instrument. As far as possible the energy scale calibration measurements were implemented as described in this international standard.

Au, Ag and Cu in form of 99.8% pure metal foils have to be used as reference samples. Certified Cu, Ag and Au foils (NPL reference metal samples SCAA90, 419) were utilized for the measurements presented here. For an XPS instrument with monochromatic $Alk_{\alpha}$ excitation the ISO 15472 ***[15]*** assigns the reference binding energies of the Au4f7/2, Ag3d5/2 and Cu2p3/2 peak to 83.96 eV, 368.21 eV and 932.62 eV, respectively. The sample have to be cleaned by ion sputtering.

Additionally the standard ISO 15472 ***[15]*** gives a firmly procedure for the determination of the measured peak energies. The ISO standard 15472 ***[15]*** stipulates to estimate the peak energy by a parabolic fit to the uppermost 20% of the intensity. Since this part of the ISO did not fit to the Quantum 2000, this issue was investigated in detail. As seen by tab. 1, only at low pass energies it is possible to record spectra with low energy grid point intervals. So only for the pass energy of 11.75 eV the peak position can be estimated by a parabolic fit. It is a time consuming task since the evaluation cannot be done within Multipak, the software package usually used for data processing. The spectral data were exported to a spread sheet program to apply a parabolic fit. The peak positions of 7 in-dependend measurements of the Au4f7/2, Ag3d5/2 and Cu2p3/2 peaks at 11.75 eV pass energy were evaluated using both Multipak fit routines and peak positions estimation by parabolic fits. Multipak fit routines give slightly lower peak positions than the parabolic fittings. The mean deviations between parabolic fit and Multipak fit routine is 26 meV for the Au4f7/2 peak, 12 meV for the Ag3d5/2 peak and 7 meV for the Cu2p3/2 peak. Deviations in this order of magnitude due to the peak position determination method are reported by Powell ***[10]***. The deviations due to the peak position determination algorithm are small compared to the achievable long time stability of the energy scale calibration. Therefore the peak position determination of the Au4f7/2, Ag3d5/2 and Cu2p3/2 signal was done by using Multipak fit routines.

The standard ISO 15472 ***[15]*** describes two different procedures. The initial procedure characterizes the XPS instrument. During the initial procedure the binding energy of Au4f7/2, Ag3d5/2 and Cu2p3/2, respectively, is measured for 7 times. These measurements determine the binding energies with high precision. Additionally they give the standard deviations of the measurements. And, hopefully, they validate the linearity of the binding energy scale. If the initial procedure shows an insufficient precision of the energy scale calibration, the instrument has to be reworked. If it is sufficient, one can proceed with subsequent calibration measurements, which control the long time stability of the energy scale. These are measurements of the Au4f7/2 and Ag3d5/2 binding energy at a regular time interval.

# Long Time Record of the Energy Scale Calibration

After one year of iterative readjustments of the instrument and initial measurements, the energy scale calibration was in an adequate condition to start with the subsequent calibration measurements. Fig. 4 shows only small deviation between the true peak binding energies as defined by the standard ISO 15472***[15]*** and the measured peak energies for the Au4f7/2, Ag3d5/2 and Cu2p3/2 signal as function of analyser pass energy. The curves are relatively flat, which means that the binding energies are nearly independent from pass energy. As mentioned above, by the subsequent calibration measurements the energy scale calibration was only checked for the pass energy settings of 11.75 eV, 46.95 eV and 93.9 eV measuring the Au4f7/2, Ag3d5/2 and Cu2p3/2 signal. Due to our experience of the first year we decided to operate the instrument with a binding energy precision or tolerance limit of ± 0.3 eV for 11.75 eV and 46.95 eV pass energy. For 93.9 eV pass energy the value is ± 0.4 eV. In extension to the standard ISO 15472***[15]*** we decided to measure the Ag3d5/2 signal, too, during the subsequent calibration measurements. This way not only the energy scale calibration but also the energy scale linearity is verified during the subsequent calibration measurements.

Fig. 5b shows the long time record of the energy calibration for the 11.75 eV pass energy setting. This is the pass energy usually used for high precision binding energy measurements. At the x-axis the dates of the measurements are written. The y-axis shows the difference

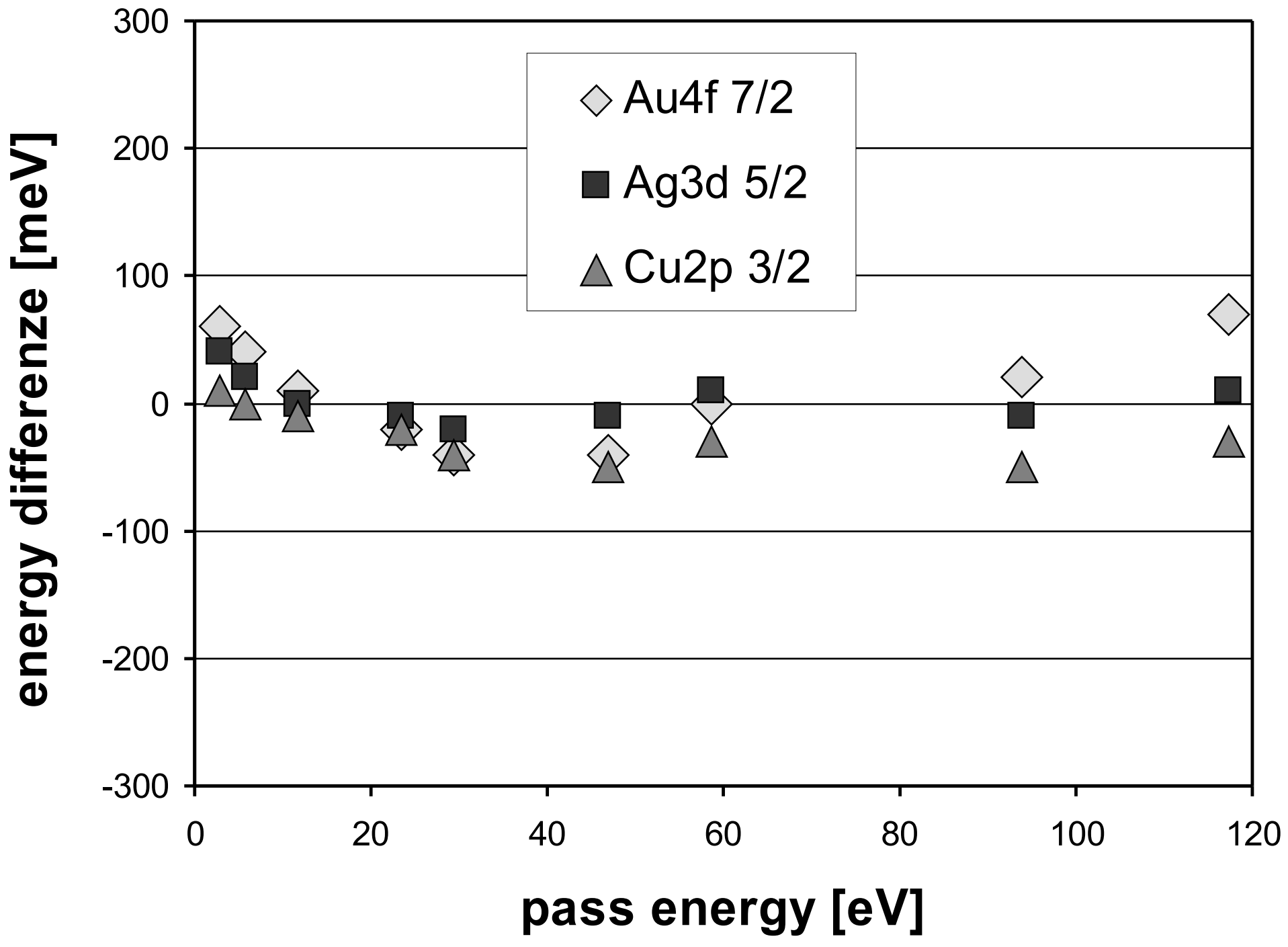

Fig. 4: energy difference between true peak binding energy and measured peak binding energy of the Au4f7/2, Ag3d5/2 and Cu2p3/2 peak measured at all pass energies
data from the year 2002

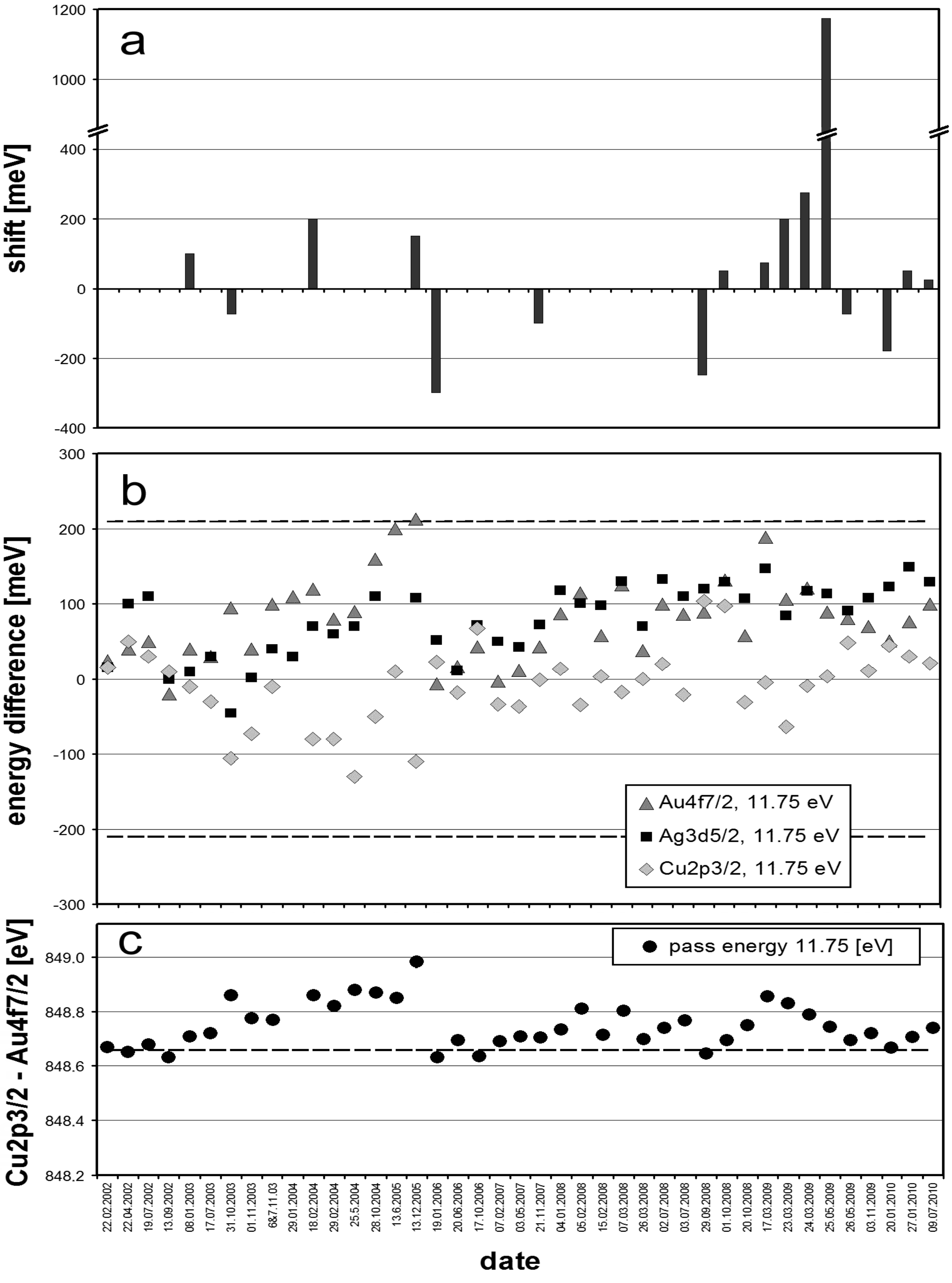

Fig. 5: data are plotted as function of measurement date, date scale is by event not linear with date, data from the year 2002 up to 2010 are shown
part a: work function shift
part b: energy difference between true peak binding energy and measured peak binding energy of the Au4f7/2, Ag3d5/2 and Cu2p3/2 peak measured at a pass energy of 11.75 eV
part c: energy difference between measured Cu2p3/2 and Au4f7/2 peak between true peak binding energy and measured peak binding energy for the Au4f7/2, Ag3d5/2 and Cu2p3/2 signals. The uncertainty $U_{95}$ ***[15]*** is ± 85 meV for the 11.75 eV pass energy setting. For clarity reasons this uncertainty is not depicted in the plot. The warning limits at ± 210 meV are plotted by dotted lines. Subsequent calibration measurements were done on a regularly time scale. If not impeded by the work load in our laboratory energy scale calibration measurements were done at the latest after 4 months. And obviously, measurements were done after elaborated system maintenance or service actions, which figure out to change the energy scale calibration. From the beginning up to Dec. 2005 the energy difference between the Au4f7/2 and Cu2p3/2 signal increases. At that time the problem was solved by a readjustment of the instrument electronics.

In general, the measurements at 46.95 eV and 93.9 eV pass energy settings point out similar results.

In fig. 5a the binding energy shift corrections of the binding energy scale are plotted. If the shift is positive, the shift was added to binding energy scale so that it was shifted to higher values and vice versa. The binding energy shift correction were calculated by minimizing the deviations between true binding energy and measured binding energy for the Au4f7/2, Ag3d5/2 and Cu2p3/2 signals for the pass energy settings of 11.75 eV, 46.95 eV and 93.9 eV. These shifts were applied to the measured binding energy data before plotting the energy differences between true peak binding energy and measured peak binding energy, which are shown in fig. 5b. Due to the discrete behaviour of digital to analogue converters the correction was a multiple of the minimal step width, which is 25 meV for our Quantum 2000.

Binding energy shifts can be explained by the deposition of material during sputter depth profiling or contaminations due to outgasing samples, which changes the work function of the energy analyser, or by slow drifts of the instrument electronics, for instance. As seen by the data, in the second period more binding energy shift corrections were applied. On one hand that may be due to an ageing of the instruments electronics. We guess, that the large corrections between March and June 2009 are caused by instruments electronic problems. Such high shifts up to ~1200 meV are not expected for work function changes. But even the replacements of parts of the analyser electronics did not reveal the origin of this problem and we were able to apply only the shift. On the other hand the type of samples, which were analysed in our laboratory, changed drastically over a period of 8 years. In the first years mainly front end technology samples like Si samples with metallic and dielectrical multilayers were analysed. In terms of material complexity and outgasing these samples are not critical. Over the years the field of application changed. More complex and outgasing samples containing organic material, like printed circuit boards, for instance, were analysed in our Quantum 2000. Most likely these outgasing sample were responsible for the binding energy shift corrections, which have to be applied more often in the second half of the reported period.

Obviously, the exact point in time between two consecutive calibration measurements, when this shift of the binding energy scale occurs, cannot be specified. Therefore, in daily laboratory practice, it is a good approach not only to measure the absolute binding energy value for a certain peak. The peaks binding energy of this certain peak should usefully be validated by peak position of others additionally measured elements being present in the sample. Then not the absolute binding energy of a certain element but binding energy differences between different elements can be used for interpretation of the measurements. One can proceed this way, as long as the energy difference between the Au4f7/2 and Cu2p3/2 signal is correct with regard to the chosen precision (see fig. 5c).

## Conclusions

It was demonstrated, that over a period of 8 years the Quantum 2000 XPS microprobe, instrument no. 78, was operated with a satisfying energy scale calibration within the chosen limits. All the time energy differences between certain peaks could be measured with the requested precision. If corrections of the energy scale calibration were necessary, in the majority of all cases these were work function shifts of the energy scale adding or subtracting constant values. Absolute binding energy values could not be measured reliably. For a multiple purpose analytical instrument unpredictable work function changes are unavoidable. These changes are caused by outgasing samples, for instance. This problem is not reported in literature by those people who developed the energy calibration procedures. That’s not astonishing, because Cu, Ag and Au reference samples are uncritical in themes of the discussion above.

The precision of the energy scale calibration was chosen to be ± 0.3 eV at the lower pass energies of 11.75 and 46.95 eV. For the pass energy of 93.9 eV it is ± 0.4 eV. These limits are not really ambitious. But from the retrospective point of view this choice turned out to be a good compromise concerning the time and service effort needed to keep the instrument calibrated.

## Acknowledgement

For fruitful discussions and valuable suggestions I would like to express my thanks to my colleagues and to the service of Physical Electronics, Europe, in particular to Mr. Groß, who was the responsible service engineer after the installation of our instrument.

## Literature

[1] G. Schön, *High resolution Auger electron spectroscopy of metallic copper*, J. Electron Spectrosc. Relat. Phenom. (1972/1973) 1, 377-387

[2] K.Asmai, *A precisely consistent energy calibration method for X-ray photoelectron spectroscopy*, J. Electron Spectrosc. Relat. Phenom. (1976) 9, 469-478

[3] C.J.Powell, N.E.Erickson and T.E.Madey, *Results of a joint auger/ESCA round robin sponsored by astm commitree E-42 on surface analysis : Part I. Esca results*, J. Electron Spectrosc. Relat. Phenom. (1979) 17, 361-403

[4] C.D.Wagner, *Energy Calibration of Electron Spectrometres*, in: *Applied Surface Analysis*, ed.: T.L.Barr and L.E.Davis, American Society for Testing and Materials (1980) 137-147

[5] M.T.Anthony and M.P.Seah, *XPS*: *Energy calibration of electron spectrometers. 1—An absolute, traceable energy calibration and the provision of atomic reference line energies*, Surf. Interface Anal. (1984) 6, 95-106

[6] M.T.Anthony and M.P.Seah, *XPS: Energy calibration of electron spectrometers. 2—Results of an interlaboratory comparison,* Surf. Interface Anal. (1984) 6, 107-115

[7] M.P.Seah, *Post-1989 calibration energies for X-ray photoelectron spectrometers and the 1990 Josephson constant*, Surf. Interface Anal. (1989) 14, 488

[8] C.J.Powell, *Energy calibration of X-ray photoelectron spectrometers: Results of an interlaboratory comparison to evaluate a proposed calibration procedure*, Surf. Interface Anal. (1995) 23, 121-132

[9] P.J.Cumpson, M.P.Seah and S.J.Spencer, *Simple Procedure for Precise Peak Maximum Estimation for Energy Calibration in AES and XPS,* Surf. Interface Anal. (1996) 24, 687-694

[10] C.J.Powell, *Energy Calibration of X-ray Photoelectron Spectrometers. II. Issues in Peak Location and Comparison of Methods*, Surf. Interface Anal. (1997) 25, 777-787

[11] A.C.Miller, C.J.Powell, U.Gelius and C.R.Anderson, *Energy calibration of X-ray photoelectron spectrometers. Part III: Location of the zero point on the binding-energy scale,* Surf. Interface Anal. (1998) 26, 606-614

[12] M.P.Seah, I.S.Gilmore and S.J.Spencer, *XPS: binding energy calibration of electron spectrometers 4 - assessment of effects for different x-ray sources, analyser resolutions, angles of emission and overall uncertainties,* Surf. Interface Anal. (1998) 26, 617-641

[13] M.P.Seah, I.S.Gilmore and G.Beamson, *XPS: binding energy calibration of electron spectrometers 5 - re-evaluation of the reference energies,* Surf. Interface Anal. (1998) 26, 642-649

[14] M.P.Seah, I.S.Gilmore and S.J.Spencer, *Measurement of data for and the development of an ISO standard for the energy calibration of X-ray photoelectron spectrometers*, Appl. Surf. Sci. (1999) 144-145, 178-182

[15] ISO 15472-2001, *Surface chemical analysis - X-ray photoelectron spectrometers - Calibration of energy scales* (2001)

[16] M.P.Seah, *Summary of ISO/TC 201 Standard: VII ISO 15472 : 2001—surface chemical analysis—x-ray photoelectron spectrometers—calibration of energy scales*. Surf. Interface Anal. (2001) 31, 721-723

[17] Physical Electronics Inc.. *System Specifications for the PHI Quantum 2000*, Eden Prairie, MN 55344 USA (1999)

[18] H. Iwai, R. Oiwa, P. E. Larson and M. Kudo, *Simulation of Energy Distribution for Scanning X-ray Probe*, Surf. Interface Anal. (1997) 25, 202-208

[19] Physical Electronics Inc.. *The PHI Quantum 2000: A Scanning ESCA Microprobe*, Eden Prairie, MN 55344 USA (1997)

[20] U.Scheithauer, *Quantitative Lateral Resolution of a Quantum 2000 X-ray Microprobe*, Surf. Interface Anal. (2008) 40, 706-709